# Assessing the properties of the prediction interval in random-effects meta-analysis

Peter Matrai[1,2], Tamas Koi[3,4], Zoltan Sipos[1,2], Nelli Farkas[1,2]

[1] Institute for Bioanalysis, Medical School, University of Pecs, Pecs, Hungary

[2] Institute for Translational Medicine, Medical School, University of Pecs, Pecs, Hungary

[3] Department of Stochastics, Institute of Mathematics, Budapest University of Technology and Economics, Budapest, Hungary

[4] Centre for Translational Medicine, Semmelweis University, Budapest, Hungary

**EMAIL ADDRESS**
**Peter Matrai**: peter.matrai@pte.hu
**Tamas Koi**: koitomi@math.bme.hu
**Zoltan Sipos**: sipos.zoltan@pte.hu
**Nelli Farkas**: nelli.farkas@aok.pte.hu

**AUTHOR CONTRIBUTIONS**
**Peter Matrai:** Conceptualization, Investigation, Methodology, Resources, Software, Visualization, Writing – Original Draft Preparation, Writing – Review & Editing
**Tamas Koi:** Conceptualization, Methodology, Resources, Visualization, Supervision, Writing – Original Draft Preparation, Writing – Review & Editing
**Zoltan Sipos:** Software, Writing – Review & Editing
**Nelli Farkas:** Conceptualization, Supervision, Writing – Review & Editing

**ACKNOWLEDGMENTS**
We acknowledge KIFÜ (Governmental Agency for IT Development, Hungary, https://ror.org/01s0v4q65) for giving us access to the Komondor high-performance computing system. We thank the HPC staff at KIFÜ for their technical support. We are grateful for the members of the Department of Biostatistics at the University of Veterinary Medicine Budapest for their valuable comments.

**FUNDING STATEMENT**
This study was not funded or sponsored.

**CONFLICT OF INTEREST STATEMENT**
The authors declare no conflicts of interest.

**DATA AVAILABILITY STATEMENT**
All data generated for the simulation study and all R codes are available through the link: https://mega.nz/folder/InZQEbrS#4AVpjRk5VsgGl_1BVH9yyw
All R codes used for the simulation are available on GitHub: https://github.com/peter-matrai/Sim_Matrai.git

**ABSTRACT**

Random effects meta-analysis is a widely applied methodology to synthetize research findings of studies in a specific scientific question. Besides estimating the mean effect, an important aim of the meta-analysis is to summarize the heterogeneity, i.e. the variation in the underlying effects caused by the differences in study circumstances. The prediction interval is frequently used for this purpose: a 95% prediction interval contains the true effect of a similar new study in 95% of the cases when it is constructed, or in other words, it covers 95% of the true effects distribution on average. In this article, after providing a clear mathematical background, we present an extensive simulation investigating the performance of all frequentist prediction interval methods published to date. The work focuses on the distribution of the coverage probabilities and how these distributions change depending on the amount of heterogeneity and the number of involved studies. Although the single requirement that a prediction interval has to fulfill is to keep a nominal coverage probability on average, we demonstrate why the distribution of coverages cannot be disregarded, and that for small number of studies no reliable conclusion can be drawn from the prediction interval. We argue that assessing only the mean coverage can easily lead to misunderstanding and misinterpretation. The length of the intervals and the robustness of the methods concerning non-normality of the true effects are also investigated.

1. INTRODUCTION

Meta-analysis (MA) is a widely used scientific methodology that includes the process of systematically searching and locating the quantitative evidence in a specific research question and integrating them using statistical methods. The most frequently applied meta-analysis model is the random-effects (RE) model, which assumes that the true effects in the distinct study populations are not equal, but they can be characterized by a common underlying distribution. Deviations between the true effects is generally referred to as heterogeneity and it can be attributed to differences in study settings and populations.

Fitting a random effects model includes the point and interval estimation of the mean parameter of the true effects and their variance, the heterogeneity parameter. The conventional report of a frequentist meta-analysis output focuses on the mean parameter, its confidence interval (CI) and the corresponding p-value. The problem with this practice is that the mean parameter and its CI gives an estimate only for the mean of the true effects and in the presence of heterogeneity there might be study populations where the effect size is crucially different [1, 2, 3, 4]. Focusing only on the mean parameter hides these differences and can lead to overconfident, oversimplified and potentially misleading conclusions. The heterogeneity in a random-effects model is conventionally assessed by the Q-test, the $I^2$ value and the $\tau^2$ statistic, however the interpretation of these measures is difficult.

In 2009 Higgins et al. [5] proposed a prediction interval (PI) for the random effects meta-analysis, which is a natural, straightforward way to summarize and report the heterogeneity. The 95% PI is an interval that covers 95% of the true effects distribution on average, or phrased differently: it contains the true effect of a hypothetical, similar new study in 95% of the cases when it is constructed. This interval has the advantage that it is on the same scale as the effect measure and contributes to a more complete summary of a meta-analysis. In the past years many authors have argued that the PI should be routinely part of a RE meta-analysis report [2, 3, 4, 18, 25], and indeed, the PI became default or at least optional setting in many meta-analysis softwares.

We found five simulation studies that aim to investigate the coverage performance of any PI estimator [1, 6, 7, 8, 9], these assess the PI performance based on the probability that the random interval contains a newly generated true effect. In their pioneering work, Brannick et al. [1] argue that the PI is not always a satisfactory method for a given research question as its target is solely the mean coverage and they investigate other intervals, e.g. the tolerance interval which have broader and more ambitious targets than just the mean coverage. Our study analyzes the PI from different aspects, we also go beyond the mean coverage probability by investigating also the distribution of coverages. We will show that investigating the distribution of the coverage probabilities reveals important aspects of the PI, which remains hidden if we assess only the mean coverage. We also introduce and investigate some other criteria to assess the performance of the PI estimators which may be important for future research to construct better and more meaningful intervals in meta-analysis.

In this article we present a comprehensive simulation that we conducted to assess the performance of all frequentist PI methods that we could find in the literature. We investigate the Higgins-Thompson-Spiegelhalter (HTS) PI [5] along with two newly proposed PI estimators, the one proposed by Wang and Lee [10] and a parametric bootstrap method proposed by Nagashima et al [6]. We also investigate how sensitive these estimators are for the case when the random effects distribution departs from normal. It is a common criticism of the RE model that the assumption of normally distributed random effects is not justified [5, 11, 12, 13], as no other simulation investigated it, we show how robust these PI estimators are if this assumption is violated.

The structure of this article is the following. In section 2 we review the random effects model, the mathematical context of the meta-analytical PI. In section 3 we give a formal definition for the PI and its assessed properties, and we give a brief description of the investigated methods and their implementation in the frequently used MA softwares. In Section 4 we show an application of the various PI methods for a published meta-analysis and we deal with the implications and concerns that arise during their interpretation. In section 5 we describe our simulation methods, in section 6 we present the results. In the discussion in section 7 we synthetize our findings and express our perception about the PI based on our results and the literature. In section 8 we give a short conclusion and write about possible further research directions.

## 2. RANDOM EFFECTS MODEL

The random effects meta-analysis model uses a probability distribution, so-called random effects to model the differences in the circumstances of the published studies. There are often differences in the involved study populations (e.g. age, education, health status), they are often conducted in various countries with different cultures and health care system, and often the studied exposure and the follow up times are meaningfully different.

In the mathematical model of the conventional two-step random effect meta-analysis, it is assumed that the true (theoretical, or population) effects of the involved studies are independent and identically distributed drawings from an unknown distribution [14]. In the conventional case, normality is assumed and the main goals of the analysis are to estimate the expected value and the variance of this distribution [15]. The published, observed outcomes are erroneous versions of the study-specific true effects. More exactly, it is assumed that a normally distributed random measurement error is added to the study-specific true effect. Let K denote the number of studies involved in the meta-analysis and let $\widehat{\theta}_k$, $\theta_k$ and $\varepsilon_k$ denote the observed outcome, the true effect and the random error in the k-th study, respectively (k=1, 2, ..., K). Then, according to the basic model discussed above

$$\widehat{\theta}_k = \theta_k + \varepsilon_k, \qquad (1)$$

$$\theta_k \sim N(\mu, \tau^2),$$

$$\varepsilon_k \sim N(0, \sigma^2{}_k),$$

where $\mu$ and $\tau^2$ denote the expected value and the variance of the true-effect distribution, respectively, $\sigma^2{}_k$ denotes the variance of the study specific sampling error $\varepsilon_k$, and $\theta_1, ..., \theta_K, \varepsilon_1, ..., \varepsilon_K$ are independent.

If $\tau^2 = 0$, the model above is called common effect model (or fixed effect model); in this case the assumption is made that each study estimates the same true effect and the observed effects differ only because of within study measurement error. It is very rare in social or medical sciences that the studies conducted in the same research question are so similar that the common effect model is plausible [5]. Even if $\widehat{\tau^2}$, the estimated between study variance parameter is 0 for a collection of studies, it is reasonable to assume in this case that we underestimate this variance and some heterogeneity is still present in the given scientific field.

The conventional random effects model specified under (1) makes two normality assumptions. The first one is that the observed effects are unbiased, normally distributed estimates of the unknown true effects. This assumption is well-founded for studies with reasonable sample size, as the central limit theorem (CLT) or for certain effect measures the maximum likelihood theory guarantees at least asymptotic normality, which gives ground to model the observed effects this way. [16] The other normality assumption represented by the second line of (1) states that the true effects are also normally distributed with the same mean and variance parameter. This assumption is often criticized as we can only argue with the CLT on this level of the model if we are willing to accept that the unexplained between study variation is the sum of many independent factors [11, 12, 13]. To examine how the PI methods perform if the true effects distribution departs from normality, we also simulated scenarios where the true effects have skewed, bimodal or uniform distribution.

### 2.1. Parameter estimation in the random effects model

The conventional aim of meta-analysis is to estimate $\mu$ and $\tau^2$. An unbiased estimate of $\mu$ can be given by a weighted mean: $\widehat{\mu} = \frac{\sum_k \widehat{\theta}_k w_k}{\sum_k w_k}$. Weights are conventionally chosen as $w_k = 1/(\sigma^2{}_k + \tau^2)$, because the variance of this estimator, $\text{Var}(\widehat{\mu}) = 1/\sum_k w_k$ provides the uniformly minimum variance estimator for $\mu$ [17]. Note that in the formula for weights, neither $\sigma^2{}_k$, nor $\tau^2$ are known quantities.

The standard approach treats $\sigma^2_k$-s as fixed and known constants and replaces them by $\widehat{\sigma^2_k}$, the within study variance estimates. This approach, just as the within study normality, is justifiable when the sample sizes are appropriately large [18]. There are numerous methods in the literature for the estimation of the $\tau^2$ parameter, a good description of these can be found in Veroniki et al. (2015) [19]. We used two frequently applied $\tau^2$ estimators to construct PI, a method of moments based estimator suggested by DerSimonian and Laird (DL) [14] and the restricted maximum likelihood (REML) estimator [20]. Hartung and Knapp [21] and Sidik and Jonkman [22] independently proposed an alternative variance estimator for the $\mu$ parameter, denoted in the following by $\widehat{\text{Var}}_{HKSJ}(\hat{\mu})$, that takes into account that $w_k$ weights are not known but estimated quantities and they use this estimator with t distribution to construct a confidence interval for $\mu$. $\widehat{\text{Var}}_{HKSJ}(\hat{\mu})$ is usually larger than $\widehat{\text{Var}}(\hat{\mu})$, but in some rare cases it can be the opposite way.

Heterogeneity is frequently assessed by the Q test introduced by Cochrane [23] testing the null hypothesis of homogeneity ($H_0: \tau^2 = 0$), and by the I² statistic or the magnitude of $\widehat{\tau^2}$. The I² statistic was proposed by Higgins and Thompson [24], it can be interpreted as the proportion of total variance in the effect estimates that can be attributed to the between study variance.

## 3. PREDICTION INTERVAL

When the between study parameter estimate, $\widehat{\tau^2}$, has a magnitude that is relevant in the specific research question, it means that $\mu$, the average effect and its confidence interval does not represent well the studied quantity, as in this case, based on the involved studies, there is a wide range of true effects that we can expect depending on unknown study conditions [13]. These plausible true effects can be meaningfully different from the overall mean, but it is not reflected in the CI. One way to handle this situation is to make an attempt to explain this large heterogeneity by forming more homogenous subgroups or fit a meta-regression model. Many times such an attempt is not successful and the researcher faces substantial between study variance at the end of the analysis. In this case, an important part of the report of the MA is to present this variance in the underlying true effects in an understandable, clear way. Reporting only the $\widehat{\tau^2}$ statistic or the I² value is not sufficient as these measures are hard to interpret and they are not on the same scale as the effect measure [2]. Higgins and his coauthors proposed a prediction interval in their 2009 paper [5], an interval that contains the true effect of a new study that is conducted in similar conditions in 95% of the cases when such an interval is constructed. It is possible to construct the PI with other than 95% probability, but we are approaching the PI mainly from a medical prospective, where the 95% is the conventional case. In the recent years, many authors in the medical field suggested that this interval should be presented on the conventional meta-analysis report [2, 3, 4, 18, 25] and several commonly used MA software and packages offers to present the PI on the forest plot as a default or optional setting (see section 3.4.).

### 3.1. Formal definition of the prediction interval

We proceed to use the notation introduced in section 2, stating that we have K studies that can be modelled by the data generating process represented by (1). Let F() denote the cumulative distribution function of the true effects distribution. The inputs of the meta-analysis are realizations from the random variables $\widehat{\Theta}_k$ along with the corresponding within study variances $\widehat{\sigma_k^2}$ (k= 1, 2, … , K). Let D denote the input variables, i.e., the collection of the random variables $\widehat{\Theta}_k, \widehat{\sigma_k^2}$. Based on a realization of D, the applied PI method outputs a prediction interval, let L(D) and U(D) denote the lower and upper PI boundaries. Note that if a realization of D is given, the PI is a deterministic function of the given realization. However, L(D) and U(D) are random variables, as the input data D can be considered as a 2K dimensional vector of random variables. The length of the interval can be defined as

$$L = U(D) - L(D) \qquad (2)$$

and the covered probability as

$$C = F(U(D)) - F(L(D)). \qquad (3)$$

**Definition 1.** In random effects meta-analysis the random interval (L(D), U(D)) with E[C] = 1 – α is called prediction interval with expected coverage probability 1 – α.

Let $\Theta_{New}$ denote the true effect in a new study, i.e., a random variable having the same distribution as $\Theta_k$-s and which is independent of D. The tower rule implies that (see Appendix for proof)

$$E[C] = P[\Theta_{New} \in (\,L(D), U(D)\,)]. \qquad (4)$$

The above equality makes clear that E[C] can be approximated by performing several times the following simulation procedure: simulating the input data D from model (1), calculating the interval (L(D), U(D)), simulating independently an additional true effect $\Theta_{New}$, then checking if $\Theta_{New} \in (\,L(D), U(D)\,)$. If the number of simulations is large, the relative frequency of the cases when the calculated interval contains the newly generated true effect will be close to E[C]. Each previous studies that investigated any PI method in the random effects meta-analysis assessed the coverage performance only by approximating E[C] based on equation (4) using the above described simulation approach [1, 6, 7, 8, 9]. Although the only requirement of a PI concerns only E[C], any researcher who understands the correct interpretation of this interval and uses the PI in his or her MA will be interested not just in E[C], but also in the distribution of C. Therefore we assessed the coverage performance of the PI methods not just based on E[C], but also based on the distribution of C defined by equation (3).

Additionally, we investigated Median[C], and E[|C - 0.95|]. To gain more insight into the behavior of the investigated intervals, we also analyzed the length of the PI. To be able to compare the observed interval length to a reasonable ideal length, we defined $L_T$, the theoretical length as the difference between the 0.975 and 0.025 quantiles of the distribution of $\Theta_k$. We assessed L by comparing it to this theoretical length and investigated $E\left[\frac{L}{L_T}\right]$ and $E\left[\left|\frac{L-L_T}{L_T}\right|\right]$.

### 3.2. Methods to construct the prediction interval

#### 3.2.1. The Higgins – Thompson – Spiegelhalter (HTS) method

Higgins, Thompson and Spiegelhalter were the first in their 2009 article to introduce the prediction interval in the context of the frequentist random effects meta-analysis [5]. They construct their PI estimator assuming $\hat{\mu} \sim N(\mu, \text{Var}(\hat{\mu}))$ and $\Theta_{New} \sim N(\mu, \tau^2)$ and the independence of $\Theta_{New}$ and $\hat{\mu}$. These infer $\Theta_{New} - \hat{\mu} \sim N(0, \tau^2 + \text{Var}(\hat{\mu}))$. They argue, that as $\tau^2$ has to be estimated and it impacts both parts of the variance of $\Theta_{New} - \hat{\mu}$, the t-distribution with K−2 degrees of freedom should be used to construct the PI. Assuming $\frac{\Theta_{New} - \hat{\mu}}{\sqrt{\hat{\tau}^2 + \widehat{\text{Var}}(\hat{\mu})}} \sim t_{K-2}$, they propose the (1 − α) expected coverage level PI as

$$\hat{\mu} \pm t_{K-2}^{\alpha}\sqrt{\hat{\tau}^2 + \widehat{\text{Var}}(\hat{\mu})} \quad (5)$$

where $t_{K-2}^{\alpha}$ is the (1−α/2) quantile of the t-distribution with K−2 degrees of freedom.

Note that using the $t_{K-2}$ distribution has no convincing theoretical basis, Viechtbauer for example offers the standard normal distribution to construct the PI as the default method in his software [26]. Partlett and Riley [7] also uses modified versions of (5) in their work. We also tested different versions of this HTS-type interval in our simulation, using the

- DL method [14] to estimate $\tau^2$ denoted by HTS-DL ($t_{k-2}$),
- REML estimator for $\tau^2$ [17] denoted by HTS-REML ($t_{k-2}$),
- HKSJ estimator for the variance of $\hat{\mu}$ [21, 22] denoted by HTS-HKSJ ($t_{k-2}$),
- $t_{k-1}$ distribution denoted by HTS-DL ($t_{k-1}$),
- standard normal distribution denoted by HTS-DL (z).

To all the HTS-type methods that are constructed with t distribution we will refer simply as HTS (t). See section 3.4. for a collection of available methods in the frequently used MA softwares.

#### 3.2.2. The parametric bootstrap method

Nagashima, Noma and Furukawa published a bootstrap PI construction method [6], which is a parametric method because it utilizes heavily the initial assumption of the normality of the random effects. They show that based on this assumption, $\theta_{New} \sim \bar{\mu} + Z\tau - t_{K-1}\sqrt{Var_{HKSJ}(\bar{\mu})}$, where $\bar{\mu}$ is an estimate for $\mu$ using the theoretical weights assuming $\sigma^2_k$ and $\tau^2$ are known, Z is a random variable with standard normal distribution, $t_{K-1}$ is random variable with t distribution with K-1 degrees of freedom, and $\sqrt{Var_{HKSJ}(\bar{\mu})}$ is the theoretical standard error of $\bar{\mu}$ defined by Hartung and Knapp [21] and Sidik and Jonkman [22]. As a next step they define a plug-in estimator for $\theta_{New}$ and give an approximate predictive distribution for $\widehat{\theta}_{New}$ as

$$\widehat{\theta}_{New} \sim \hat{\mu} + Z\hat{\tau}_{UDL} - t_{K-1}\sqrt{\widehat{Var}_{HKSJ}(\hat{\mu})}, \tag{6}$$

where $\hat{\mu}$ and $\sqrt{\widehat{Var}_{HKSJ}(\hat{\mu})}$ are computed as $\sigma^2_k$ and $\tau^2$ are not known, but estimated quantities, and $\hat{\tau}_{UDL}$ is the untruncated DerSimonian and Laird estimator for $\tau$ [14]. As the formula for $\widehat{\theta}_{New}$ contains 3 random variables (Z, $t_{K-1}$ and $\hat{\tau}_{UDL}$), they generate B independent random realizations from these distributions and compute $\widehat{\theta}_{New}$ for each realization based on (6). Generating realizations from the distribution of $\hat{\tau}_{UDL}$ is difficult, it is based on the distribution of the Q statistics described by Biggerstaff and Jackson [27]. For the boundaries of the (1 − α) expected coverage level PI, they simply take the α/2 and (1−α/2) quantiles of the B realization of $\widehat{\theta}_{New}$.

### 3.2.3. The ensemble method

Wang and Lee [10] used the background idea of Louis [28] and constructed a PI estimator by modifying the observed effects so that the empirical distribution of $\widehat{\theta}_k$ have a variance that equals to $\tau^2$ asymptotically. They define $\widehat{\theta}^*_k = \hat{\mu} + \sqrt{\frac{\hat{\tau}^2}{\hat{\tau}^2+\widehat{\sigma^2_k}}}(\widehat{\theta}_k - \hat{\mu})$ and show that $Var(\widehat{\theta}^*_k) \approx \tau^2$ for large K. As a next step they simply use the α/2 and (1−α/2) quantiles of $\widehat{\theta}^*_k$ to create a PI with (1 − α) expected coverage level. Note that if $\hat{\tau}^2 = 0$, this method yields a single number, $\hat{\mu}$ as PI, but the HTS type intervals in this case give an interval that is at least as wide as the CI.

### 3.3. Relationship between the prediction interval and the tolerance interval

Although it is rarely used in practice, it is possible to define the tolerance interval for the random effects meta-analysis. The tolerance interval aims to contain at least a specific proportion (β) of the true effects distribution in 1 − α of the times it is constructed.

**Definition 2** In random effects meta-analysis the random interval (L(D), U(D)) is a β-content tolerance interval if P [C ≥ β] = 1 – α, where β is the interval content (coverage proportion) and 1 – α is the confidence level.

Brannick et al. [1] modified certain non-meta-analysis-specific tolerance interval construction methods and tested by simulation whether the resulting intervals fulfill the definition above with β = 0.8 and α = 0.05. They argue convincingly that many times the tolerance interval suits better to the given research question as the prediction interval aims to contain a certain proportion of the true effects only *on average*, however the tolerance interval aims to contain *at least this proportion* with some predefined confidence level [29].

### 3.4. PI estimators in common MA softwares

Since its introduction for the random effects meta-analysis in the frequentist framework in 2009 [5], the PI became optional part in most of the commonly used MA softwares and program packages. The available and the default methods show notable variety, we investigated the newest versions of those MA softwares that we think are commonly used and made a table showing the default and the other available PI methods in these softwares (Table 1). The most common ones are the different HTS-type intervals, the parametric bootstrap method is only available in the meta R package [30]. The ensemble method is not implemented in any of these softwares, but it is available freely in a supplementary spreadsheet file published along with their article [10].

### 4. EXAMPLE

We chose a real, highly cited meta-analysis conducted on the medical field to illustrate the place of the prediction interval in the interpretation of the summary results and also the arising concerns. Thase and his co-authors published a meta-analysis in 2016 investigating the short term (6-8 weeks) efficacy of the active agent vortioxetine in adult patients suffering from major depressive disorder (MDD) [31]. Their analysis included 11 randomized, double-blind, placebo controlled trials investigating the effect of fix doses of vortioxetine of 5, 10, 15 and 20 mg. Not each trial had study arms with each dose, we chose their 10 mg analysis because it included the most trials, 7. The primary endpoint was the mean change from baseline relative to placebo in the MADRS total score. The MADRS is a depression rating scale ranging between 0 and 60, with higher points indicating more severe depression, a reduction of 2 points in the scale compared to placebo is generally considered clinically meaningful [31]. Based on the study level data that they represent on their Figure 2A, we reconstructed their 10 mg dose analysis and also computed the 95% prediction interval with each method described in section 3.2. (Figure 1). Based on the random effects model, the mean effect with its 95% CI shows a clear advantage of vortioxetine over placebo: $\hat{\mu}$ = -3.57, CI: -4.97; -2.17. However, there is substantial heterogeneity: the $I^2$ measure is 65%, the observed effects of the individual studies range between -7.18 and -0.78, and there is one study (NCT00839423) which has a CI with no overlapping part with CI-s of 3 other studies. In this case the $\hat{\mu}$ and its CI mask these important differences between the individual studies, however, reporting the PI reveals them. The PI computed with any method is much wider than the CI, the widest is the parametric bootstrap PI (-8.1; 0.94). The HTS (t) intervals are just a slightly shorter and these also cross the null effect line, only the HTS-DL (z) and the ensemble method give a PI that do not pass the null effect line, only these 2 lie entirely on the beneficial part of the scale. These PI-s therefore suggest that in some populations, that are similar to the ones represented in these 7 studies, a null effect or even a harmful effect of 10 mg vortioxetine is possible. How confident can we be, that these conclusions are reliable considering that they are based only on 7 studies? Which method should we trust? We will address these questions in the discussion.

## 5. METHODS

Our meta-analysis input simulation approach is similar to the effect size simulation method of Bakbergenuly et al. [32]. We investigate a continuous variable and determine the mean difference (MD) as an effect measure between 2 groups, an experimental (E) and a control (C) group. First, we fix K, the number of studies involved in the meta-analysis, as K in {3, 4, 5, 7, 10, 15, 20, 30, 100}; k represents the index of a given study in the meta-analysis, k=(1, 2, … ,K). Then we determine the total sample size in a given study, $N_k$ by one of 2 ways. The sample sizes are either equal in all included studies, in this case $N_1 = N_2 = … = N_K$ in {30, 50, 100, 200, 500, 1000, 2000}, or if they are mixed, the vector (50, 100, 500) is repeated periodically to $N_K$, namely $N_1 = 50$, $N_2 = 100$, $N_3 = 500$, $N_4 = 50$ … $N_K$. For the sample sizes in the 2 groups, $N_{E,k}$ and $N_{C,k}$, we simply halve the total sample size: $N_{E,k} = N_{C,k} = N_k/2$. We fix the population variance (V) of the continuous variable in the 2 groups as $V_{E,k} = V_{C,k} = 10$. This does not lead to loss of generality as we will control the standard error of the mean difference by controlling the sample size. As a next step, we calculate the theoretical variance of the mean difference, the squared standard error or within study variance, $\sigma^2_k$, and simulate its observed value in the sample, $\widehat{\sigma^2}_k$. For these, we first generate the sample variance, $\widehat{V}$, in the 2 groups as a scaled random realization from a $\chi^2$ distribution with appropriate degrees of freedom: $\widehat{V}_{E,k} \sim \chi^2(df = N_{E,k} - 1) * \frac{V_{E,k}}{N_{E,k}-1}$ and $\widehat{V}_{C,k} \sim \chi^2(df = N_{C,k} - 1) * \frac{V_{C,k}}{N_{C,k}-1}$ and compute the observed squared standard error as $\widehat{\sigma^2_k} = \frac{\widehat{V}_{E,k}}{N_{E,k}} + \frac{\widehat{V}_{C,k}}{N_{C,k}}$ and its theoretical version as $\sigma^2_k = \frac{V_{E,k}}{N_{E,k}} + \frac{V_{C,k}}{N_{C,k}}$.

To simulate the true effects, $\theta_k$, and their observed value, $\hat{\theta}_k$, we first fix $\tau^2$, the variance of the true effects as $\tau^2$ in {0.1, 0.2, 0.3, 0.5, 1, 2, 5}, and $\mu$, the expected value of the true effects as $\mu = 0$. To simulate the true effects, we used the normal distribution, and 5 other distributions to investigate how robust the methods are for the deviation from normality. We simulated $\theta_k$ from slightly, moderately and highly skewed normal distributions with skewness coefficients of 0.5, 0.75 and 0.99, respectively. We also simulated true effects from a bimodal distribution mimicking the case where there is an underlying factor that divides the studies into 2 subgroups. We generated random numbers from a bimodal distribution as a mixture of 2 normals using the R package EnvStats v. 2.8.1., [33] and from the skewed normal distributions using the sn package v. 2.1.1. [34]. As an extremity, we also investigated the case where the true effects follow the uniform distribution. The density function of these distributions is visualized in the Supplementary Material [35]. As each investigated simulation factors are fully crossed, the number of the total investigated scenarios are: (N) * (K) * ($\tau^2$) * (true effects distributions) = 8 * 9 * 7 * 6 = 3024.

We chose the mean difference as the effect measure in our simulation because it is a frequently used effect size in the medical field, but we believe that the conclusions are generalizable to other effect sizes (standardized mean difference, hazard ratio, odds ratio, risk ratio, etc.) because previous simulations show that the main factors influencing PI performance are the number of involved studies and the amount of heterogeneity [6, 7], regardless of the effect measure.

For each of these scenarios we simulated 5000 meta-analysis realizations and computed the PI with each investigated methods, with the exception of the parametric bootstrap method, which we evaluated for 1000 realizations for each scenario. The parametric bootstrap method is a highly computation intensive method requiring many bootstrap samples, this parameter can be set by the B parameter of the pima function in the pimeta R package. The default setting of the function is B=25000, Nagashima et al. prepared their simulation with B=5000 repetitions [6]. We also used the B=5000 setting in our simulation. We used the high-performance computing system of the Hungarian Governmental Agency for IT Development (KIFÜ) to evaluate the parametric bootstrap method and a simple personal computer for the other methods. The simulation was programmed in R v.4.1.3. [36].

We measured the heterogeneity of the scenarios by 2 ways. We calculated the ratio $v = \frac{\tau^2}{\overline{\sigma^2}}$ following Partlett and Riley [7], where $\overline{\sigma^2}$ is the mean within study variance. We also considered the mean of the observed $I^2$ measures for each scenario. We chose our simulation parameters so that both v and $I^2$ cover a wide range of values. Our investigated scenarios cover the range of 0.08 - 250 of v values and the 7% - 99% range of the mean $I^2$ values.

### 5.1. Investigated performance measures

For each simulation scenario and PI method, we calculated the empirical versions of the theoretical quantities described at the end of section 3.1. Thus, we examined and visualized the empirical distribution of the coverage probabilities on histograms. We computed the mean of the coverage probabilities, the median coverage probability and also the mean absolute difference from 95% coverage approximating E[C], Median[C] and E[|C - 0.95|], respectively. To investigate the length of the PI-s, we examined the mean of the observed length relative to their true length approximating $\mathrm{E}\left[\frac{L}{L_T}\right]$, and also the mean relative distance from the true length approximating the quantity $\mathrm{E}\left[\left|\frac{L-L_T}{L_T}\right|\right]$, which we will refer to as normalized mean absolute error on the figures and in the results section.

## 6. RESULTS

All of our simulation results are available in the Supplementary Material [35]. This is a compact, easy to use html file which contains each performance measure plots for each of our simulation scenarios. We included figures for two scenarios in the main text of this article, a low heterogeneity scenario (Figures 2-4: N=100, $\tau^2$=0.2, $I^2$=33%, v=0.5) and a high heterogeneity scenario (Figures 5-7: N=100, $\tau^2$=1, $I^2$=71%, v=2.5). In both of these presented scenarios the random effects are normally distributed. In sections 6.1.- 6.6. we describe results for the case where true effects follow the normal distribution, and in section 6.7. we give a description about the cases where the true effects have skewed, bimodal or uniform distribution.

### 6.1. Histogram of coverage probabilities

The histograms of coverage probabilities show the same pattern for each method: for any given sample size and $\tau^2$ combination the histograms are very left-skewed unless the number of involved studies is extremely large (K=100). As there are more and more studies in the meta-analysis, the coverage distribution gets less and less left-skewed and starts cumulating on the nominal level of 95%. This pattern is true in general for all scenarios but for smaller heterogeneity (Figures 2-3) this process is slower and even for large number of studies the distribution remains fairly left-skewed. For higher heterogeneity scenarios, (Figures 5-6) as the number of studies increases, the coverage distributions reach a concentrated, less skewed distribution faster, compared to the smaller heterogeneity scenarios. For very large number of studies (K=100) the distribution shows an ideal, fairly symmetric shape that is concentrated on the nominal coverage provided the heterogeneity is large (Figures 5-6, d, h, l). For less heterogeneous scenarios the coverage distribution remains a bit left skewed even if the study number is very large (Figures 2-3, d, h, l).

### 6.2. Mean coverage probability

The HTS (t) methods give a very high, nearly 100% mean coverage for K=3 studies, and then as the number of studies increases, their mean coverage drops below the nominal level. For low heterogeneity these methods only retain the nominal coverage when the involved studies are very large (Figure 4a). When the heterogeneity is larger, (Figure 7a), this drop is less dramatic and they yield a mean coverage close to the nominal level even for K=15 studies. The HTS-DL (z) and the ensemble method give a very low mean coverage for the lower heterogeneity scenarios and they only retain mean coverages close to the nominal level when both heterogeneity and the number of studies is large. The mean coverage of the parametric bootstrap method remains close to the nominal level irrespective of heterogeneity and study number, and this method also produces mean coverages close to 95% for larger heterogeneity and larger study number scenarios.

### 6.3. Median coverage

The median coverage probabilities differ in some important aspects from the mean coverages (Figures 4b and 7b). The HTS (t) and the parametric bootstrap method yield in general a higher than 95% median coverage, and they only retain 95% median coverage for scenarios with very large study number. The parametric bootstrap method tends to give higher median coverages than the HTS-type methods that are constructed with t distribution, especially for the lower heterogeneity scenarios. The HTS-DL (z) method and the ensemble method give a median coverage well below the nominal level for the lower heterogeneity, lower study number scenarios and as the study number increases, they approach the 95% level from below 95%. For the higher heterogeneity scenarios the HTS-DL (z) and the ensemble method can retain the 95% median coverage even if the study number is not very large (K ≥ 10).

### 6.4. Mean absolute difference from 95% coverage

The HTS (t) methods give about a 5% mean absolute difference for the K=3 study number scenarios (Figures 4c and 7c). For the higher study number scenarios this increases, especially for the lower heterogeneity cases (Figure 4c), and for scenarios with very large study number (K=100) it gets closer to 0 again, but even for K=30, they give about a 10% mean absolute difference from the nominal coverage. For higher heterogeneity scenarios (Figure 7c) this initial increase is smaller and the mean absolute difference remains smaller than it is for the small heterogeneity cases. The HTS-DL (z) method gives a higher mean absolute difference than the HTS (t) intervals for the lower study number scenarios, just as the ensemble method. For higher heterogeneity and study number scenarios this difference gets smaller and when v ≥ 5 and K ≥ 15 each method yield very similar mean absolute difference. The parametric bootstrap method gives the smallest mean absolute difference of all methods, except the very low study number scenarios (K = 3-5).

### 6.5. Mean observed length relative to true length

The HTS-DL ($t_{k-2}$), HTS-HKSJ ($t_{k-2}$), HTS-DL ($t_{k-1}$) and the parametric bootstrap method produce too long intervals on average compared to the expected true length especially when the number of studies is low. As the study number increases their mean observed length get closer and closer to the true length (Figures 4d and 7d). For the lower heterogeneity scenarios (Figure 4d) the parametric bootstrap method tends to give even longer intervals on average than the HTS (t) methods. The HTS-DL (z) and the ensemble method yield a mean observed interval close to the true interval, except the heterogeneity is very low ($I^2$ < 22%) and the number of involved studies is also low (K < 7).

### 6.6. Normalized mean absolute error

The parametric bootstrap and the HTS (t) methods give a relatively high mean absolute error if the study number is low (K < 7) and with the increasing study number this error decreases (Figures 4e and 7e). The HTS-DL (z) and the ensemble method give smaller mean absolute error for these low study number scenarios compared to the other investigated methods. If the heterogeneity is high (Figure 7e) all methods produce smaller normalized mean absolute error compared to the low heterogeneity scenarios (Figure 4e).

### 6.7. Impact of non-normality of the true effects

When the true effects distribution departs from normal, the histograms are similar to the case when it follows the normal distribution. The only exception is where the random effects are drawn from the uniform distribution. In this case a spike remains at >99% coverage for all methods irrespective of the other circumstances and the histograms do not approach a concentrated, symmetric form as the number of studies increases.

The mean, median and mean absolute difference coverages are very similar to the normal distribution case described above when the true effects distribution is skewed or bimodal, but for the uniform scenario, methods give higher mean coverage and mean absolute difference compared to the normal case and nearly 100% median coverage regardless of any other parameters. Methods tend to give slightly higher mean observed length and normalized mean absolute error when true effects have non-normal distribution, especially when their distribution is uniform.

## 7. DISCUSSION

Our aim with this study was to conduct an in-depth analysis of the performance of the PI in the frequentist random effects meta-analysis, investigate its usefulness and limitations and help researchers to understand its correct interpretation and point out the possible misinterpretations.

We argue that the most complete and correct understanding about the behavior and interpretation of this interval can be achieved by investigating the distribution of the coverage probabilities. To our best knowledge, our study is the first to investigate the distribution of coverage probabilities, and how these distributions change depending on factors like heterogeneity, the number of involved studies and the true effects distribution. Looking at the histograms either on Figures 2-3 and 5-6 or in our Supplementary Material [35], it is clear that the distribution of coverages cannot be disregarded and why assessing only the mean coverage leads to oversimplification and possible misunderstanding.

Our simulation results confirm the findings of Nagashima et al. [6] regarding that the parametric bootstrap method achieves the nominal level of mean coverage even if the number of involved studies are very small, e.g. K = 3-5. The HTS-type and the ensemble methods are frequently unable to maintain the 95% mean coverage, especially when the number of studies and the heterogeneity is low (Figure 4a).

The problem gets clear if we look at the histogram of coverages, for example the K=5 scenario of the parametric bootstrap method on Figure 2(e) or 5(e), we can see that in 60-70% of the cases, depending on the heterogeneity, the PI computed by this method covers more than 99% of the true effects distribution, it gives a coverage close to 95% very rarely, and it produces eventually a mean coverage of 95% only because the few cases of very low, 20-60% coverages balance out the overwhelming majority of coverages exceeding even 99%. Unfortunately, this pattern remains fairly stable when the number of involved studies are larger (e. g. Figures 2 and 5, f and g), and reaches a close to ideal form only for scenarios with large study number. On Figure 5h, which shows the distribution for large heterogeneity and K=100 studies, we can see that the parametric bootstrap method gives 95% coverage not only on average, but we can expect that all the individual coverages will be close to 95% as well. For lower heterogeneity but same study number (K=100), the distribution remains less concentrated on the nominal level (Figure 2h).

This is the basic pattern and the general problem with all PI methods, but we wanted to describe it with the parametric bootstrap method because this retains the nominal coverage on average regardless of the amount of heterogeneity or study number, and the other methods are frequently unable to maintain it not even on average.

The above described findings have very serious implications for any researcher who is conducting a meta-analysis and wants to use the PI to summarize the results and wants to draw conclusions based on it. The common interpretation of the PI is that this is an interval of 'plausible' [7] or 'expected' [2, 19] range of true effects in similar studies or simply that this is a 'range of true effects in future studies' [10]. We think that researchers or readers of the meta-analysis might misinterpret the above or similar definitions and believe that this is an interval that covers 95% of the true effects distribution or an interval that covers the true treatment effect in a future similar study with 95% probability [8]. The distribution of coverages clearly show that unfortunately, this interpretation is only valid on average and cannot be generalized for one certain case of meta-analysis, unless the number of involved studies are extremely large. This has serious consequences for the researcher or for the reader because he or she will see only one prediction interval each time he or she conducts or reads a meta-analysis. The very left skewed coverage distributions imply that with high probability he or she will have an interval that is too conservative with too much coverage and will not take comfort in the notion that because of the low probabilities of too little coverages eventually the mean coverage of all possible scenarios is optimal. We think that he or she would be more interested in the coverage probability for that single, certain interval that the given method produces for that certain collection of studies he or she has at hand, rather than just an abstract mean coverage. The former can be best investigated if we show the whole distribution and not just the mean, because even if the mean coverage is optimal, it might be the case that the coverage of almost every single interval is too far from this optimal level.

The very left skewed distribution of PI coverages gives an explanation why so many meta-analyses with statistically significant mean parameter give a PI that contains the null effect or even the opposite value of the mean effect. IntHout et al. re-evaluated 479 statistically significant MA from the Cochrane Database published between 2009–2013 and found that 72% of these has a 95% PI that contains the null effect and 20% of the calculated 95% PIs contain the opposite effect, suggesting that the effect in similar populations could be even the opposite, i.e. harmful [2]. Siemens et al. conducted a similar study analyzing MA studies in the 2010-2019 period searching in multiple databases and identified 261 MA with significant mean effect. They report that in 75% of these, the 95% PI contains the null effect and 38% contains the opposite effect [3]. The explanation for these large proportion of PIs crossing the no effect line or even containing clinically relevant harmful effect values is the low number of involved studies. Our analysis shows the left skewed distribution of coverages, which, with high probability results in way much coverage than intended and essentially too wide intervals in the majority of MAs. We showed that this is especially true for MAs with low number of studies. IntHout et al. re-analyzed studies with a median study number of K=4 [37], and the median study number in Siemens et al. is K=6 [3]. In light of our analysis, their findings are not surprising but rather anticipated.

The MA presented in section 4 concludes to have a significant mean effect of 10 mg vortioxetine based on 7 studies. 5 of the investigated methods yield a PI crossing the no effect line and the remaining 2 has an upper bound very close to it. Our simulation shows that for a scenario very similar to this particular example (N = 100, $I^2$ = 71%, K = 7, see Supplementary Material [35]), the probability of a 95% PI with extremely too much coverage (>99%) is 20-50% depending on the chosen method, the mean coverage is 94% for the bootstrap method, and about 90% for the other methods or even lower (83-85% for the HTS-DL(z) and ensemble methods). The median coverage of the bootstrap method and the HTS (t) methods for this scenario is 98-99%, meaning that in 50% of the cases, these methods give a 95% PI that has coverage at least 98-99%. The HTS-DL(z) and ensemble methods give a 93-94% median coverage for this scenario. Our length analysis reveals that on average, the bootstrap method and the HTS (t) methods give a 30% wider interval than the true length, the ensemble method gives on average about 10% wider interval compared to the true length, and the HTS-DL(z) method retains the true length on average, for this scenario. These findings imply that for this particular MA the bootstrap method and the HTS type methods computed with t distribution probably yield a too wide PI with more coverage than intended, giving an explanation why they suggest a potential null or even harmful effect for similar study populations. The PI length produced by the ensemble method and the HTS-DL(z) method in this case is probably closer to the true length than the other methods, this is what both the mean observed length and the normalized mean absolute error show for this scenario. Unfortunately, these methods give a too low median coverage and cannot maintain the nominal mean coverage, meaning that they yield a PI with too low coverage more often than the other methods.

Some authors argue that the PI is helpful for the power calculation of a new trial [18, 2]. We think that whenever the PI is used in the planning of a new study, it is important to keep in mind that the coverage probability of that particular PI can seriously differ from the nominal level, as we have shown it. This is a clear example where the analyst is interested in the coverage probability of that single PI he or she wants to use to the sample size or power calculation of a next study and the mean coverage is unimportant, therefore we discourage the use of the PI for this purpose.

There are different thresholds proposed in the literature for the number of involved studies under which the use of PI is not advisable. Partlett and Riley concludes that if the heterogeneity is large enough ($I^2$ > 30%) and the study sizes are balanced, the HTS type methods perform well if K ≥ 5 [7]. Nagashima et al. states that their bootstrap method is valid even if the number of included studies are very small and performs well even when K = 3 [6]. Siemens et al. claims that reporting the PI for K ≥ 4 helps decision making [3]. The current version of the Cochrane Handbook (6.4) encourages researchers to use the PI when K ≥ 10 and if there is no sign of funnel plot asymmetry [25]. Based on our findings, we agree with the current advice of the Cochrane Handbook. We do not think it is advisable to calculate and present the PI for a meta-analysis with K < 10, unless the researcher is aware of and satisfied with the concept of the mean coverage and explicitly states it in the report. If the researcher is more interested in the coverage probability for that particular collection of studies, we showed that the current methods are unable to produce such a reliable interval unless the number of involved studies are very large. We agree with Brannick et al. that without sufficient information we cannot anticipate accurate estimates and that in these cases it is better not to report any PI [1]. Cox [38] and Higgins [5] argue that for small number of studies it is worth considering to report only the individual study results and not to report any summary statistics. Our analysis clearly shows that this is particularly true for the PI.

## 8. CONCLUSIONS

Although the only condition that a prediction interval has to fulfill is to retain a nominal coverage probability on average, we think that researchers and readers of meta-analyses would rather have an interval that has a near nominal coverage every time it is constructed and they might misinterpret the PI this way. We argue that showing the distribution of coverage probabilities is the best way to understand the correct interpretation of this interval. Inspecting these distributions reveals that for small number of studies no reliable conclusion can be drawn from the PI even if we know that it fulfills the above criterion and keeps the nominal mean coverage.

The tolerance interval has more ambitious goals than the PI, it intends to cover at least a pre-specified portion of the true effects distribution with a certain confidence level. Further research is needed how such an interval can be constructed in the random effects meta-analysis or how the currently existing PI methods can be altered to fulfill more strict conditions than the simple mean coverage criterion.

**HIGHLIGHTS**

**What is already known**
- The routine report of a random-effects meta-analysis presents the estimated mean parameter and its confidence interval, however, these do not reveal the heterogeneity, i.e. the differences in the underlying effects of individual studies.
- The prediction interval (PI) in random-effects meta-analysis is a useful tool to assess the heterogeneity, it intends to contain the true effect of a new similar study with a pre-specified probability when it is constructed.
- The Higgins-Thompson-Spiegelhalter PI method cannot keep the nominal mean coverage for cases when the number of studies is small, but the parametric bootstrap method can maintain appropriate mean coverage even if there are only 3 involved studies.

**What is new**
- Although by definition the only requirement a PI method has to fulfill is to keep the nominal mean coverage, to obtain a clear understanding about the behavior of this interval, it is also important to assess the distribution of coverages.
- If the number of involved studies is not large enough, the distribution of coverage probabilities is skewed, meaning that with high probability the coverage probability is close to 1, however, the mean coverage can still be close to the nominal level.
- The distribution of coverages reveals that even if the nominal mean coverage is maintained, the interpretation of a single, published PI: 'an interval that covers the true treatment effect in a future similar study with 95% probability' is not valid, unless there are extremely large number of available individual studies. In other words, if one considers a published PI with nominal level 0.95, it will not be true that approximately 95% of the upcoming individual studies will be within the published interval.

**Potential impact for *Research Synthesis Methods* readers**
- Even if the nominal mean coverage is approximately achieved, researchers using the PI should be cautious with the interpretation of the PI.
- Further research is needed how new methods can be developed or how the currently existing PI methods can be altered to fulfill more strict conditions than the simple mean coverage criterion.

**REFERENCES**

1       Brannick, M. T., French, K. A., Rothstein, H. R., Kiselica, A. M., & Apostoloski, N. (2021). Capturing the underlying distribution in meta-analysis: Credibility and tolerance intervals. *Research synthesis methods*, *12*(3), 264–290. https://doi.org/10.1002/jrsm.1479

2       IntHout, J., Ioannidis, J. P., Rovers, M. M., & Goeman, J. J. (2016). Plea for routinely presenting prediction intervals in meta-analysis. *BMJ open*, *6*(7), e010247. https://doi.org/10.1136/bmjopen-2015-010247

3       Siemens, W., Meerpohl, J. J., Rohe, M. S., Buroh, S., Schwarzer, G., & Becker, G. (2022). Reevaluation of statistically significant meta-analyses in advanced cancer patients using the Hartung-Knapp method and prediction intervals-A methodological study. *Research synthesis methods*, *13*(3), 330–341. https://doi.org/10.1002/jrsm.1543

4       Riley, R. D., Higgins, J. P., & Deeks, J. J. (2011). Interpretation of random effects meta-analyses. *BMJ (Clinical research ed.)*, *342*, d549. https://doi.org/10.1136/bmj.d549

5       Higgins, J. P., Thompson, S. G., & Spiegelhalter, D. J. (2009). A re-evaluation of random-effects meta-analysis. *Journal of the Royal Statistical Society. Series A, (Statistics in Society)*, *172*(1), 137–159. https://doi.org/10.1111/j.1467-985X.2008.00552.x

6       Nagashima, K., Noma, H., & Furukawa, T. A. (2019). Prediction intervals for random-effects meta-analysis: A confidence distribution approach. *Statistical methods in medical research*, *28*(6), 1689–1702. https://doi.org/10.1177/0962280218773520

7       Partlett, C., & Riley, R. D. (2017). Random effects meta-analysis: Coverage performance of 95% confidence and prediction intervals following REML estimation. Statistics in medicine, 36(2), 301–317. https://doi.org/10.1002/sim.7140

8       Hamaguchi, Y., Noma, H., Nagashima, K., Yamada, T., & Furukawa, T. A. (2021). Frequentist performances of Bayesian prediction intervals for random-effects meta-analysis. *Biometrical journal. Biometrische Zeitschrift*, *63*(2), 394–405. https://doi.org/10.1002/bimj.201900351

9       Gnambs, T., & Schroeders, U. (2024). Accuracy and precision of fixed and random effects in meta-analyses of randomized control trials for continuous outcomes. *Research synthesis methods*, *15*(1), 86–106. https://doi.org/10.1002/jrsm.1673

10      Wang, C. C., & Lee, W. C. (2019). A simple method to estimate prediction intervals and predictive distributions: Summarizing meta-analyses beyond means and confidence intervals. *Research synthesis methods*, *10*(2), 255–266. https://doi.org/10.1002/jrsm.1345

11      Baker, R., & Jackson, D. (2008). A new approach to outliers in meta-analysis. *Health care management science*, *11*(2), 121–131. https://doi.org/10.1007/s10729-007-9041-8

12      Jackson, D., & White, I. R. (2018). When should meta-analysis avoid making hidden normality assumptions?. *Biometrical journal. Biometrische Zeitschrift*, *60*(6), 1040–1058. https://doi.org/10.1002/bimj.201800071

13      Kontopantelis, E., & Reeves, D. (2012). Performance of statistical methods for meta-analysis when true study effects are non-normally distributed: A simulation study. *Statistical methods in medical research*, *21*(4), 409–426. https://doi.org/10.1177/0962280210392008

14      DerSimonian, R., & Laird, N. (1986). Meta-analysis in clinical trials. *Controlled clinical trials*, *7*(3), 177–188. https://doi.org/10.1016/0197-2456(86)90046-2

15      Whitehead, A., & Whitehead, J. (1991). A general parametric approach to the meta-analysis of randomized clinical trials. *Statistics in medicine*, *10*(11), 1665–1677. https://doi.org/10.1002/sim.4780101105

16      Brockwell, S. E., & Gordon, I. R. (2001). A comparison of statistical methods for meta-analysis. *Statistics in medicine*, *20*(6), 825–840. https://doi.org/10.1002/sim.650

17      Viechtbauer, W. (2005). Bias and Efficiency of Meta-Analytic Variance Estimators in the Random-Effects Model. Journal of Educational and Behavioral Statistics, 30(3), 261-293. https://doi.org/10.3102/10769986030003261

18	Veroniki, A. A., Jackson, D., Bender, R., Kuss, O., Langan, D., Higgins, J. P. T., Knapp, G., & Salanti, G. (2019). Methods to calculate uncertainty in the estimated overall effect size from a random-effects meta-analysis. *Research synthesis methods*, *10*(1), 23–43. https://doi.org/10.1002/jrsm.1319

19	Veroniki, A. A., Jackson, D., Viechtbauer, W., Bender, R., Bowden, J., Knapp, G., Kuss, O., Higgins, J. P., Langan, D., & Salanti, G. (2016). Methods to estimate the between-study variance and its uncertainty in meta-analysis. *Research synthesis methods*, *7*(1), 55–79. https://doi.org/10.1002/jrsm.1164

20	Raudenbush SW. (2009). Analyzing effect sizes: random-effects models. In Cooper H, Hedges LV, Valentine JC (eds.). The Handbook of Research Synthesis and Meta-Analysis (pp. 295–315). New York: Russell Sage Foundation

21	Hartung, J., & Knapp, G. (2001). A refined method for the meta-analysis of controlled clinical trials with binary outcome. *Statistics in medicine*, *20*(24), 3875–3889. https://doi.org/10.1002/sim.1009

22	Sidik, K., & Jonkman, J. N. (2002). A simple confidence interval for meta-analysis. *Statistics in medicine*, *21*(21), 3153–3159. https://doi.org/10.1002/sim.1262

23	Cochran, W. G. (1937). Problems Arising in the Analysis of a Series of Similar Experiments. Supplement to the Journal of the Royal Statistical Society, 4(1), 102–118. https://doi.org/10.2307/2984123

24	Higgins, J. P., & Thompson, S. G. (2002). Quantifying heterogeneity in a meta-analysis. *Statistics in medicine*, *21*(11), 1539–1558. https://doi.org/10.1002/sim.1186

25	Deeks JJ, Higgins JPT, Altman DG (editors). Chapter 10: Analysing data and undertaking meta-analyses. In: Higgins JPT, Thomas J, Chandler J, Cumpston M, Li T, Page MJ, Welch VA (editors). Cochrane Handbook for Systematic Reviews of Interventions version 6.4 (updated August 2023). Cochrane, 2023. Available from www.training.cochrane.org/handbook.

26	Viechtbauer W (2010). "Conducting meta-analyses in R with the metafor package." *Journal of Statistical Software*, 36(3), 1–48. doi:10.18637/jss.v036.i03.

27	Biggerstaff, B. J., & Jackson, D. (2008). The exact distribution of Cochran's heterogeneity statistic in one-way random effects meta-analysis. *Statistics in medicine*, *27*(29), 6093–6110. https://doi.org/10.1002/sim.3428

28	Louis, T. A. (1984). Estimating a Population of Parameter Values Using Bayes and Empirical Bayes Methods. Journal of the American Statistical Association, 79(386), 393–398. https://doi.org/10.1080/01621459.1984.10478062

29	Vardeman, S. B. (1992). What about the Other Intervals? The American Statistician, 46(3), 193–197. https://doi.org/10.2307/2685212

30	Balduzzi S, Rücker G, Schwarzer G (2019). "How to perform a meta-analysis with R: a practical tutorial." *Evidence-Based Mental Health*, 153–160.

31	Thase, M. E., Mahableshwarkar, A. R., Dragheim, M., Loft, H., & Vieta, E. (2016). A meta-analysis of randomized, placebo-controlled trials of vortioxetine for the treatment of major depressive disorder in adults. *European neuropsychopharmacology : the journal of the European College of Neuropsychopharmacology*, *26*(6), 979–993. https://doi.org/10.1016/j.euroneuro.2016.03.007

32	Bakbergenuly, I., Hoaglin, D. C., & Kulinskaya, E. (2020). Estimation in meta-analyses of mean difference and standardized mean difference. *Statistics in medicine*, *39*(2), 171–191. https://doi.org/10.1002/sim.8422

33	Millard SP (2013). *EnvStats: An R Package for Environmental Statistics*. Springer, New York. ISBN 978-1-4614-8455-4, https://www.springer.com.

34	Azzalini, A. (2023). The R package 'sn': The Skew-Normal and Related Distributions such as the Skew-t and the SUN (version 2.1.1). http://azzalini.stat.unipd.it/SN/,https://cran.r-project.org/package=sn

35      Matrai P., Koi T., Sipos Z., Farkas N., (2024) Assessing the properties of the prediction interval in random-effects meta-analysis. Supplementary Material. Available at: https://mega.nz/file/NjphiQZR#D-GijLzwAdWuQoNEl0Jn-opnMWbGYC8QJb78iBlfcaQ

36      R Core Team (2022). R: A language and environment for statistical computing. R Foundation for Statistical Computing, Vienna, Austria. URL https://www.R-project.org/.

37      IntHout, J., Ioannidis, J. P., Borm, G. F., & Goeman, J. J. (2015). Small studies are more heterogeneous than large ones: a meta-meta-analysis. *Journal of clinical epidemiology*, *68*(8), 860–869. https://doi.org/10.1016/j.jclinepi.2015.03.017

38      Cox, D. R. (2006) Combination of data. In Encyclopedia of Statistical Sciences, 2nd ed. (eds S. Kotz, C. B. Read, N. Balakrishnan and B. Vidakovic), pp. 1074–1081. Hoboken: Wiley.

39      Durrett, R. (2019). Probability: Theory and Examples (5th ed.). Cambridge: Cambridge University Press.

| Software | License type | Default PI method | Other available methods |
|---|---|---|---|
| R meta package v. 7.0-0 | Freeware | HTS-REML ($t_{k-2}$) | HTS-DL ($t_{k-2}$) <br> HTS-HKSJ ($t_{k-2}$) <br> Parametric bootstrap |
| R metafor package v. 4.6-0 | Freeware | HTS-REML (z) | HTS-DL (z) <br> HTS-DL ($t_{k-1}$) <br> HTS-DL ($t_{k-2}$) <br> HTS-REML ($t_{k-2}$) <br> HTS-HKSJ ($t_{k-2}$) |
| Stata v. 18 meta command | Commercial | HTS-REML ($t_{k-2}$) | HTS-DL ($t_{k-2}$) <br> HTS-HKSJ ($t_{k-2}$) |
| Stata metan routine | Commercial | HTS-DL ($t_{k-2}$) | HTS-REML ($t_{k-2}$) |
| SPSS v. 29.0.2.0. | Commercial | HTS-REML ($t_{k-2}$) | HTS-DL ($t_{k-2}$) <br> HTS-HKSJ ($t_{k-2}$) |
| Comprehensive Meta-Analysis v. 4. | Commercial | HTS-DL ($t_{k-2}$) | - |
| RevMan | Freeware | - | - |

**TABLE 1** Default and other available prediction interval estimators in common meta-analysis softwares

**Abbreviations:** HTS, Higgins – Thompson – Spiegelhalter method; REML, Restricted maximum likelihood estimation of $\tau^2$ parameter; DL, DerSimonian and Laird estimation of $\tau^2$ parameter; HKSJ, Hartung - Knapp - Sidik – Jonkman variance estimation for the µ parameter; $t_{k-2}$, method calculated with t distribution with k-2 degrees of freedom; z, method calculated with standard normal distribution; ' - ', denote that the given method is not available.

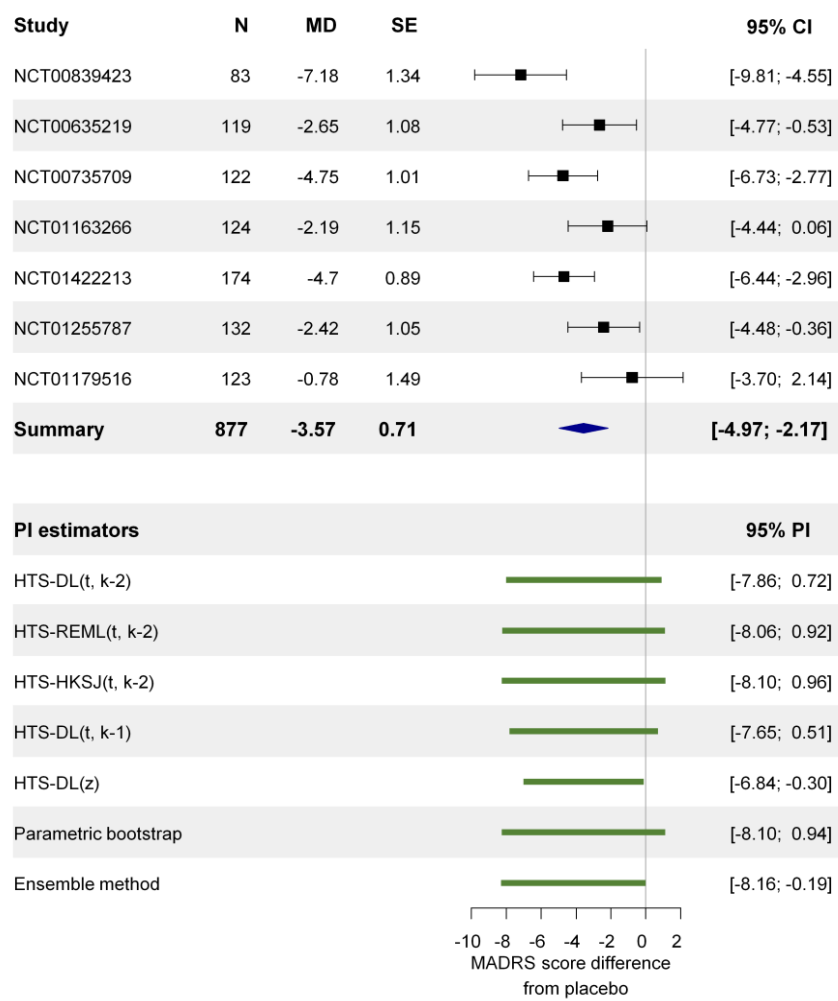

**FIGURE 1** Forest plot showing the short term efficacy of the active agent vortioxetine 10 mg in adult patients suffering from major depressive disorder based on the meta-analysis of Thase and his co-authors [31].

**Abbreviations:** N, total sample size; MD, mean difference; SE, standard error; CI, confidence interval; PI, prediction interval; MADRS, Montgomery-Åsberg Depression Rating Scale; HTS, Higgins–Thompson–Spiegelhalter method; REML, Restricted maximum likelihood estimation of $\tau^2$ parameter; DL, DerSimonian and Laird estimation of $\tau^2$ parameter; HKSJ, Hartung - Knapp - Sidik – Jonkman variance estimation for the μ parameter; $t_{k-2}$, method calculated with t distribution with k-2 degrees of freedom; z, method calculated with standard normal distribution.

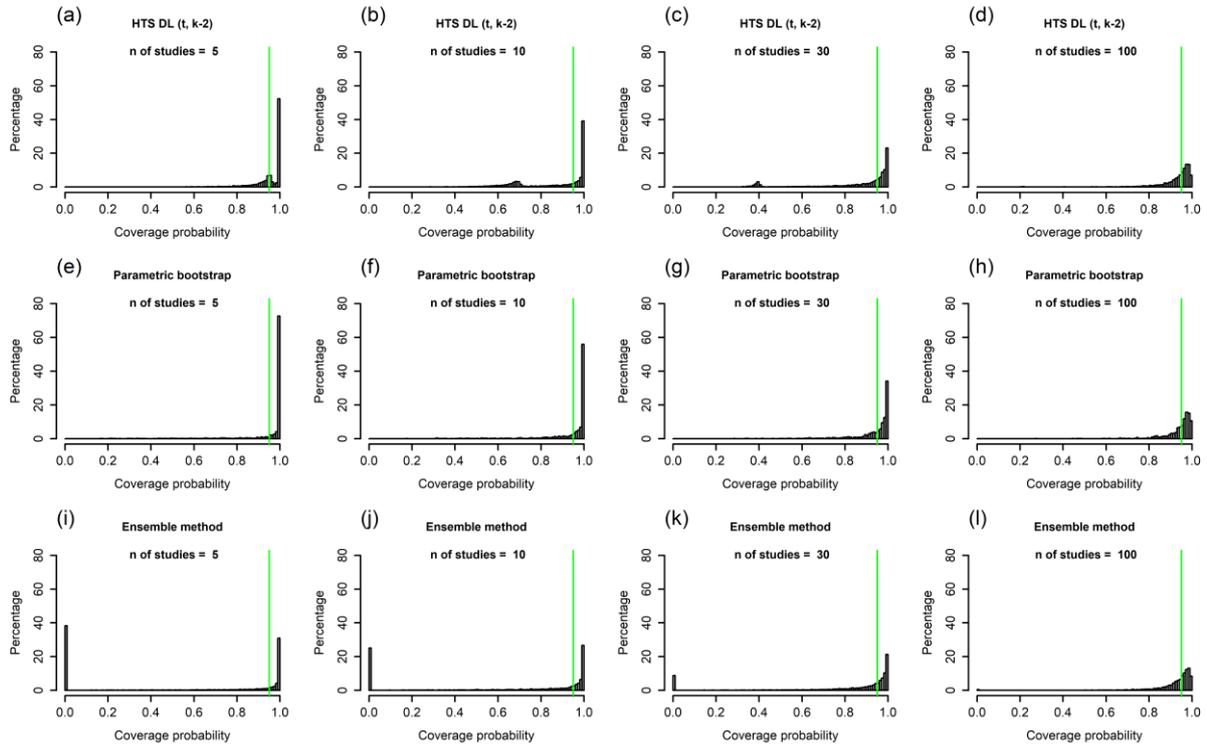

**FIGURE 2** Histograms showing the coverage probability distribution of the HTS-DL ($t_{k-2}$), the parametric bootstrap and the ensemble prediction interval methods for a low heterogeneity simulation scenario (N=100, $\tau^2$=0.2, $I^2$=33%, v=0.5). The vertical green line on the histograms indicates 95% coverage probability. The number of involved studies is constant in each column, showing the distribution for 5, 10, 30 and 100 studies. Results for the HTS-DL ($t_{k-2}$) method are represented in the first row of histograms with letters (a), (b), (c) and (d), the parametric bootstrap method is represented in the second row of histograms with letters (e), (f), (g) and (h), and the ensemble method is represented in the third row of histograms with letters (i), (j), (k) and (l).

**Abbreviations:** HTS-DL ($t_{k-2}$), Higgins – Thompson – Spiegelhalter method with the DerSimonian and Laird estimation of $\tau^2$ parameter and using the t distribution with k-2 degrees of freedom.

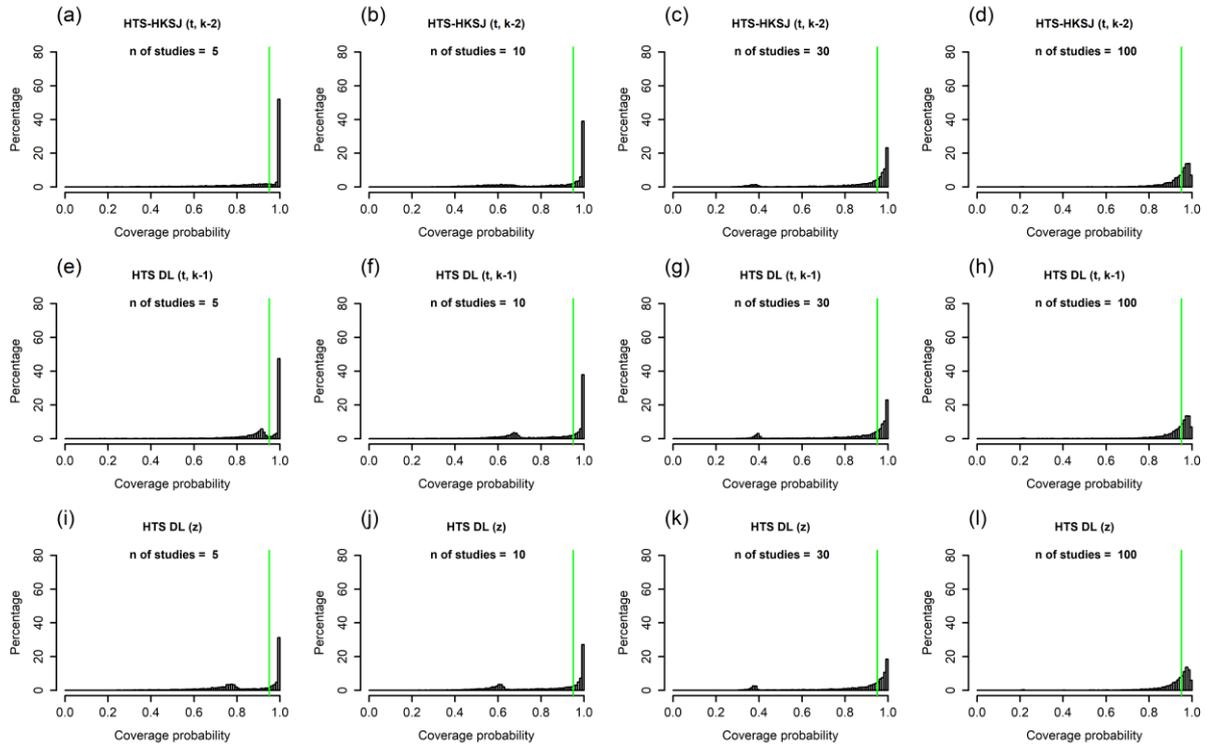

**FIGURE 3**  Histograms showing the coverage probability distribution of the HTS-HKSJ ($t_{k-2}$), the HTS-DL ($t_{k-1}$) and the HTS-DL (z) prediction interval methods for a low heterogeneity simulation scenario (N=100, $\tau^2$=0.2, $I^2$=33%, v=0.5). The vertical green line on the histograms indicates 95% coverage probability. The number of involved studies is constant in each column, showing the distribution for 5, 10, 30 and 100 studies. Results for the HTS-HKSJ ($t_{k-2}$) method are represented in the first row of histograms with letters (a), (b), (c) and (d), the HTS-DL ($t_{k-1}$) method is represented in the second row of histograms with letters (e), (f), (g) and (h), and the HTS-DL (z) method is represented in the third row of histograms with letters (i), (j), (k) and (l).

**Abbreviations:** HTS-HKSJ ($t_{k-2}$), Higgins – Thompson – Spiegelhalter method with the Hartung - Knapp - Sidik – Jonkman variance estimation for the µ parameter and using the t distribution with k-2 degrees of freedom; HTS-DL ($t_{k-1}$), Higgins – Thompson – Spiegelhalter method with the DerSimonian and Laird estimation of $\tau^2$ parameter and using the t distribution with k-1 degrees of freedom; HTS-DL (z), Higgins – Thompson – Spiegelhalter method with the DerSimonian and Laird estimation of $\tau^2$ parameter and using the standard normal distribution.

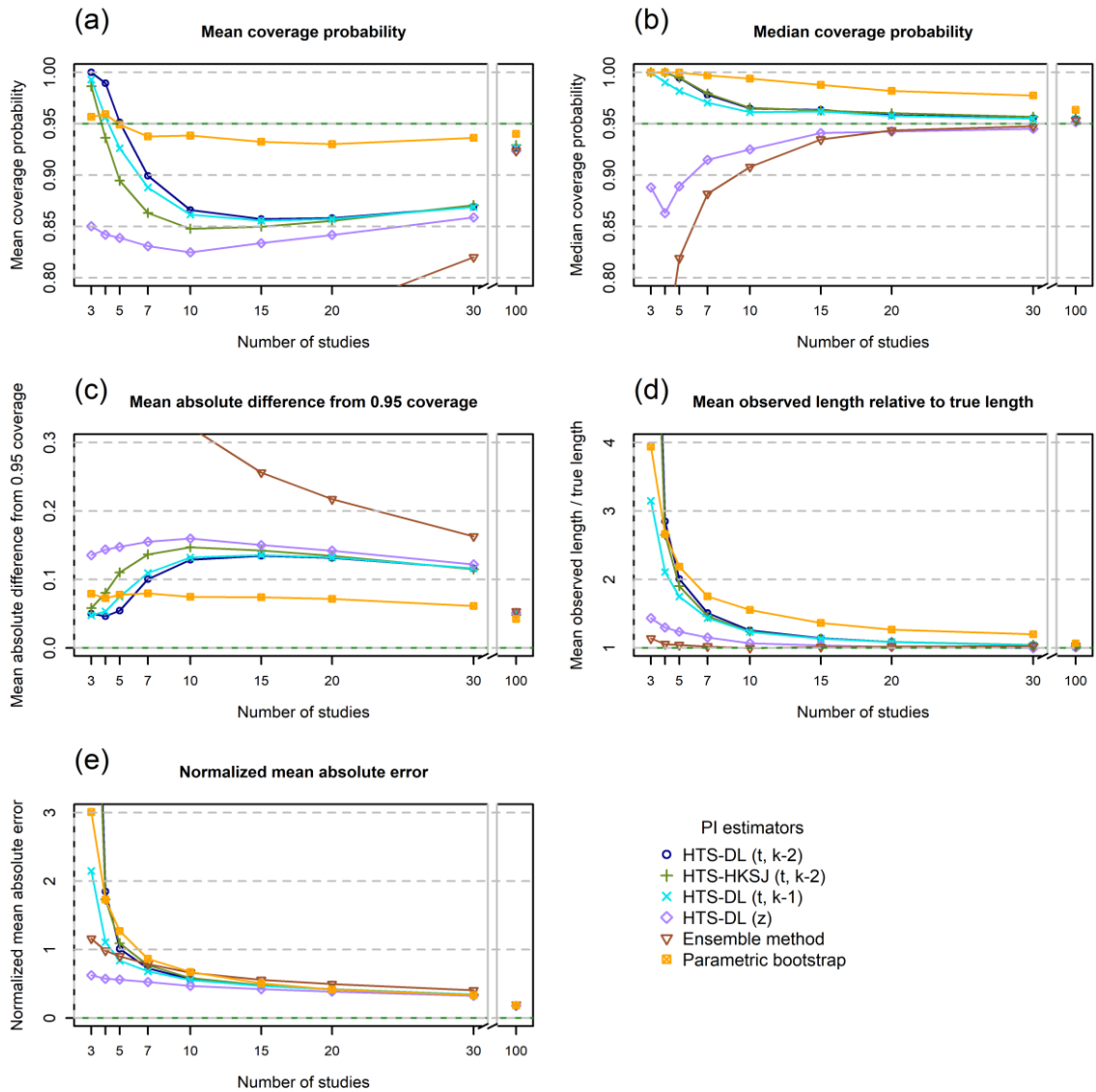

**FIGURE 4** Performance measure results of the investigated prediction interval methods as a function of the number of involved studies (horizontal axis) for a low heterogeneity simulation scenario (N=100, $\tau^2$=0.2, $I^2$=33%, v=0.5). Mean coverage probability (a), Median coverage probability (b), Mean absolute difference from 0.95 coverage (c), Mean observed length relative to true length (d), Normalized mean absolute error (e).

**Abbreviations:** HTS, Higgins – Thompson – Spiegelhalter method; DL, DerSimonian and Laird estimation of $\tau^2$ parameter; HKSJ, Hartung - Knapp - Sidik – Jonkman variance estimation for the $\mu$ parameter; $t_{k-2}$, method calculated with t distribution with k-2 degrees of freedom; z, method calculated with standard normal distribution

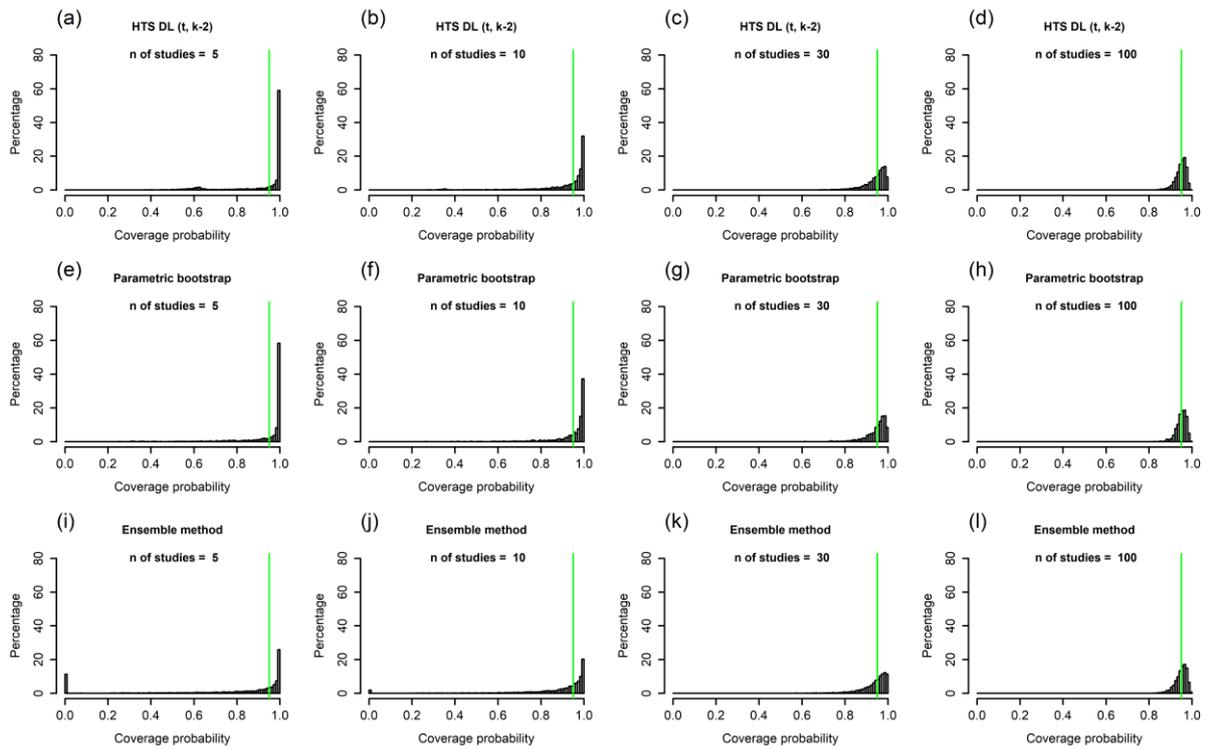

**FIGURE 5** Histograms showing the coverage probability distribution of the HTS-DL ($t_{k-2}$), the parametric bootstrap and the ensemble prediction interval methods for a high heterogeneity simulation scenario (N=100, $\tau^2$=1, $I^2$=71%, v=2.5). The vertical green line on the histograms indicates 95% coverage probability. The number of involved studies is constant in each column, showing the distribution for 5, 10, 30 and 100 studies. Results for the HTS-DL ($t_{k-2}$) method are represented in the first row of histograms with letters (a), (b), (c) and (d), the parametric bootstrap method is represented in the second row of histograms with letters (e), (f), (g) and (h), and the ensemble method is represented in the third row of histograms with letters (i), (j), (k) and (l).

**Abbreviations:** HTS-DL ($t_{k-2}$), Higgins – Thompson – Spiegelhalter method with the DerSimonian and Laird estimation of $\tau^2$ parameter and using the t distribution with k-2 degrees of freedom.

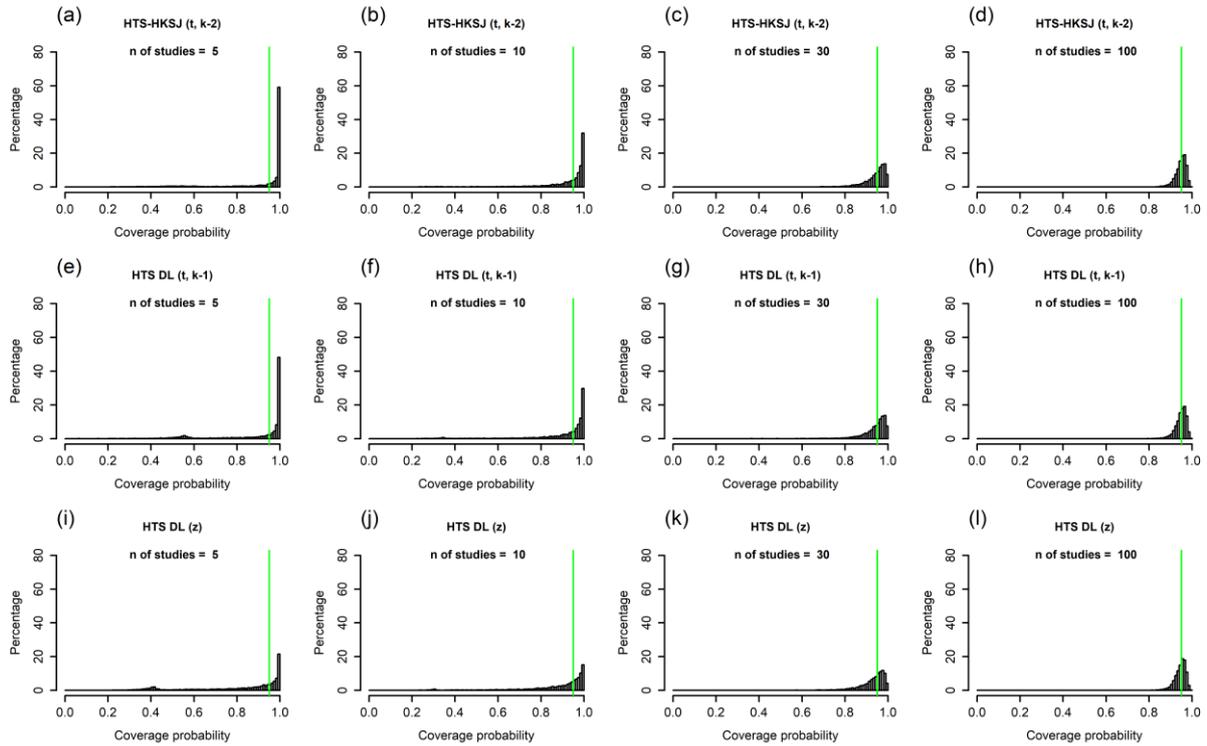

**FIGURE 6**  Histograms showing the coverage probability distribution of the HTS-HKSJ ($t_{k-2}$), the HTS-DL ($t_{k-1}$) and the HTS-DL (z) prediction interval methods for a high heterogeneity simulation scenario (N=100, $\tau^2$=1, $I^2$=71%, v=2.5). The vertical green line on the histograms indicates 95% coverage probability. The number of involved studies is constant in each column, showing the distribution for 5, 10, 30 and 100 studies. Results for the HTS-HKSJ ($t_{k-2}$) method are represented in the first row of histograms with letters (a), (b), (c) and (d), the HTS-DL ($t_{k-1}$) method is represented in the second row of histograms with letters (e), (f), (g) and (h), and the HTS-DL (z) method is represented in the third row of histograms with letters (i), (j), (k) and (l).

**Abbreviations:** HTS-HKSJ ($t_{k-2}$), Higgins – Thompson – Spiegelhalter method with the Hartung - Knapp - Sidik – Jonkman variance estimation for the µ parameter and using the t distribution with k-2 degrees of freedom; HTS-DL ($t_{k-1}$), Higgins – Thompson – Spiegelhalter method with the DerSimonian and Laird estimation of $\tau^2$ parameter and using the t distribution with k-1 degrees of freedom; HTS-DL (z), Higgins – Thompson – Spiegelhalter method with the DerSimonian and Laird estimation of $\tau^2$ parameter and using the standard normal distribution.

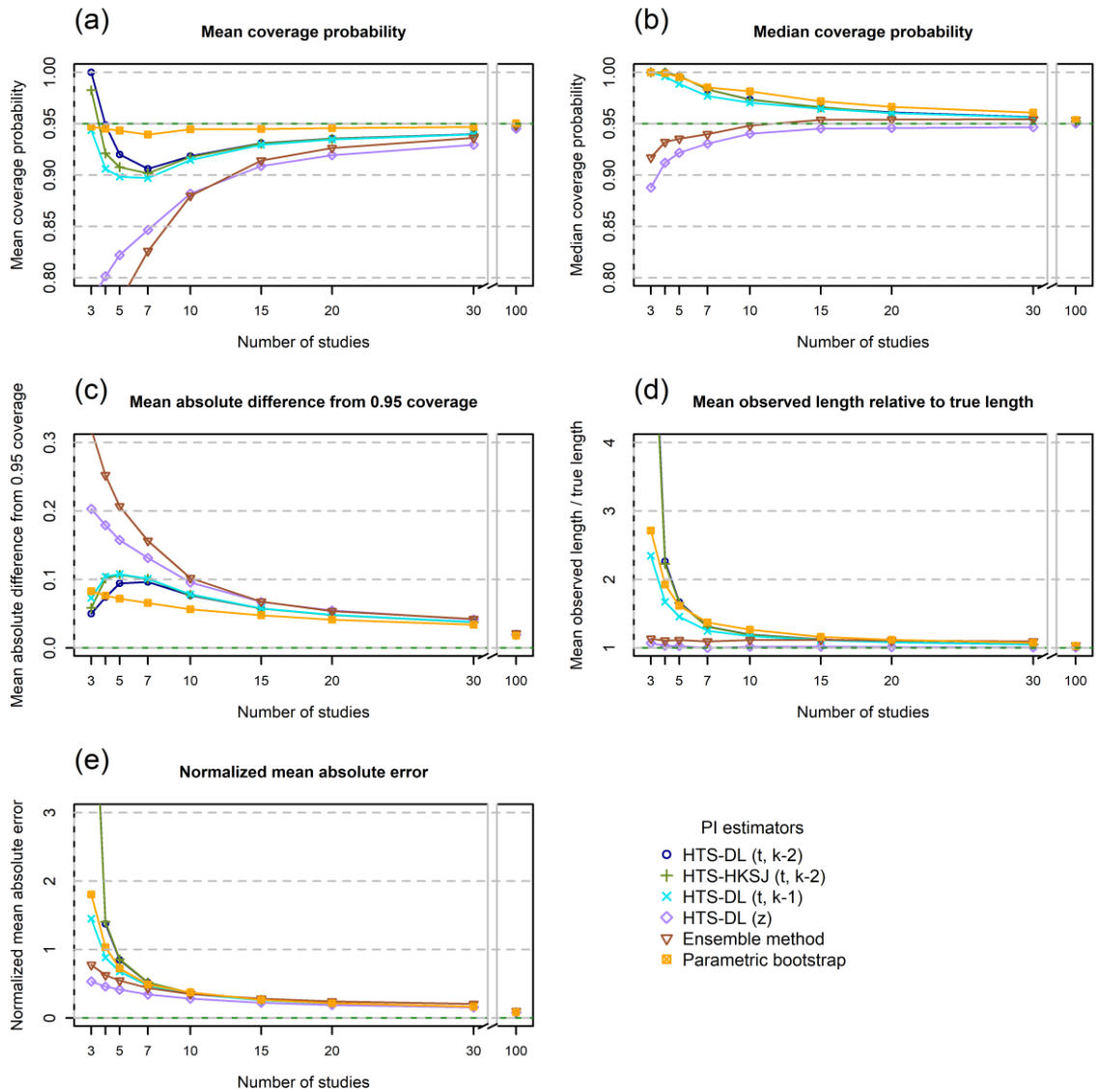

**FIGURE 7** Performance measure results of the investigated prediction interval methods as a function of the number of involved studies (horizontal axis) for a high heterogeneity simulation scenario (N=100, $\tau^2=1$, $I^2=71\%$, v=2.5). Mean coverage probability (a), Median coverage probability (b), Mean absolute difference from 0.95 coverage (c), Mean observed length relative to true length (d), Normalized mean absolute error (e).

**Abbreviations:** HTS, Higgins – Thompson – Spiegelhalter method; DL, DerSimonian and Laird estimation of $\tau^2$ parameter; HKSJ, Hartung - Knapp - Sidik – Jonkman variance estimation for the µ parameter; $t_{k-2}$, method calculated with t distribution with k-2 degrees of freedom; z, method calculated with standard normal distribution;

Appendix for the article

**Assessing the properties of the prediction interval in random-effects meta-analysis**

Peter Matrai[1,2], Tamas Koi[3,4], Zoltan Sipos[1,2], Nelli Farkas[1,2]

1  Institute for Bioanalysis, Medical School, University of Pecs, Pecs, Hungary

2  Institute for Translational Medicine, Medical School, University of Pecs, Pecs, Hungary

3  Department of Stochastics, Institute of Mathematics, Budapest University of Technology and Economics, Budapest, Hungary

4  Centre for Translational Medicine, Semmelweis University, Budapest, Hungary

**Proof that the expected value of the covered probability C, defined in section 3.1. in equation (3) fulfills equation (4)**

In this section we will prove that the expected value of the covered probability C, defined in (3) fulfills (4). For the sake of completeness, first we review important properties about the conditional expectation based on Chapter 4.1. in Durrett's textbook Probability: Theory and Examples [39].

When Y is a random variable on the probability space ($\Omega; \mathcal{F}_0; P$) with $E(|Y|) < \infty$ and $\mathcal{F}$ is a sub sigma-algebra of $\mathcal{F}_0$ then the conditional expectation $E(Y|\mathcal{F})$ is almost surely well defined. In the sequel all equalities between random variables mean almost sure equality. In particular, when $\mathcal{F}$ is the trivial sigma-algebra, i.e., $\mathcal{F} = \{\emptyset, \Omega\}$ then $E(Y|\mathcal{F})$ is equal to the conventional expected value $E(Y)$.

If both *X* and *Y* are random variables on the probability space ($\Omega; \mathcal{F}_0; P$), then the conditional random variable $E(Y|X)$ appearing in probability courses can be reformulated as $E(Y|X) = E(Y|\sigma(X))$, where $\sigma(X)$ is the sigma-algebra generated by *X*, i.e., the smallest sigma-algebra according to which *X* is measurable. It will be crucial in the upcoming proof that the probability of an event $A \in \mathcal{F}_0$ can be expressed as

$$P(A) = E(\mathbb{1}_A), \qquad (7)$$

where $\mathbb{1}_A$ is the indicator function of event *A*, i.e., the random variable on ($\Omega; \mathcal{F}_0; P$) defined by

$$\mathbb{1}_A(\omega) = \begin{cases} 1 & \omega \in A \\ 0 & \omega \notin A \end{cases} \qquad (8)$$

Note also that *P(A|X)* can be defined as

$$P(A|X) = E(\mathbb{1}_A|X). \qquad (9)$$

**Proposition 1** (Example 4.1.7. in Durrett [39]) Suppose $X$ and $Y$ are independent. Let $\varphi$ be a function with $E(|\varphi(X,Y)|) < \infty$ and let $g(x) = E(\varphi(x, Y))$. Then

$$E(\varphi(X, Y)|X) = g(X). \tag{10}$$

**Proposition 2** (Special case of formula (4.1.5) in Durrett [39], also known as tower rule). Let $X$ and $Y$ be random variables on the probability space $(\Omega; \mathcal{F}_0; P)$ with $E(|Y|) < \infty$. Then

$$E(E(Y|X)) = E(Y). \tag{11}$$

To prove (4), first observe that the independence of $\Theta_{New}$ and D, (9) and Proposition 1 imply that

$$P[\Theta_{New} \in (L(D), U(D))|D] = E[\mathbb{1}_{\Theta_{New} \in (L(D), U(D))} | D] = E[\varphi(\Theta_{New}, D) | D] =$$

$$F(U(D)) - F(L(D)) = C. \tag{12}$$

Finally, the above equality, (8), (9) and Proposition 2 imply

$$E[C] = E\left[E[\mathbb{1}_{\Theta_{New} \in (L(D), U(D))} | D]\right] = \tag{13}$$

$$E[\mathbb{1}_{\Theta_{New} \in (L(D), U(D))}] = \tag{14}$$

$$P[\Theta_{New} \in (L(D), U(D))]. \tag{15}$$

Hence, the claimed equality (4) is proved.